# Examining the Effectiveness of Chatbots in Gathering Family History Information in Comparison to the Standard In-Person Interview-Based Approach

Kieron Drumm, Vincent Tran

*School of Computing, Dublin City University, Dublin, Ireland*

One of the most common things that a genealogist is tasked with is the gathering of a person's initial family history, normally via in-person interviews or with the use of a platform such as ancestry.com, as this can provide a strong foundation upon which a genealogist may build. However, the ability to conduct these interviews can often be hindered by both geographical constraints and the technical proficiency of the interviewee, as the interviewee in these types of interviews is most often an elderly person with a lower than average level of technical proficiency. With this in mind, this study presents what we believe, based on prior research, to be the first chatbot geared entirely towards the gathering of family histories, and explores the viability of utilising such a chatbot by comparing the performance and usability of such a method with the aforementioned alternatives. With a chatbot-based approach, we show that, though the average time taken to conduct an interview may be longer than if the user had used ancestry.com or participated in an in-person interview, the number of mistakes made and the level of confusion from the user regarding the UI and process required is lower than the other two methods. Note that the final metric regarding the user's confusion is not applicable for the in-person interview sessions due to its lack of a UI. With refinement, we believe this use of a chatbot could be a valuable tool for genealogists, especially when dealing with interviewees who are based in other countries where it is not possible to conduct an in-person interview.

CCS Concepts: • **Human-centered computing** → **Empirical studies in HCI**

Additional Key Words and Phrases: chatbot, genealogy, ancestry, family history research

## 1. INTRODUCTION

The idea of knowing, recording, and preserving one's family history has gained popularity in recent years. This could be due to the potential healing aspects of ancestral storytelling, the pursuit of knowledge about where one's parents or grandparents came from and what they experienced [1], or the recording of other people's family histories for the sake of preservation. Though the terms "genealogy" and "family history research" are used somewhat interchangeably nowadays, there is a clear difference between the two, specifically concerning the depth to which the research goes [2]. Where genealogy seeks to track down key information about an individual's ancestors, such as their date and place of birth, whether or not they were married, and when and where they died, family history research seeks to go further by "unearthing supplementary information about ancestors' home, educational, working, social and political lives" [2]. For this review, both terms will be used interchangeably, but it should be noted that the type of research being referred to here is closer to the "genealogical" definition — focusing specifically on the key information associated with a person's ancestors. Genealogical research can be a pleasantly open-ended task for some, limited solely by practicalities (whether or not a record is accessible or one is willing to expend effort) [2]. Though this study aims to determine how effectively a chatbot could potentially replace an FHR in conducting interviews, this is not the only means by which genealogists and FHRs conduct their research. Many FHRs visit local libraries, archives, national centres and niche collections in order to conduct research [2]. The overall purpose of this study is to examine the effectiveness of a chatbot, or conversational AI, in carrying out such research and the comparison between a chatbot, a family history researcher, and ancestry.com's family tree builder with regard to information gathered and the accuracy of said information. See Table 1 for a direct comparison between the methods examined in this study that shows how our chatbot could be a useful tool for any genealogist when dealing with someone with a low level of technical proficiency who lives far enough away from them to rule out the possibility of an in-person interview.

Table 1. A Comparison of the Information Gathering Methods Examined in this Study.

|  | **Does it exist?** | **Technical skills required** | **Can it be done remotely?** |
|---|---|---|---|
| **In-person Interviews** | Yes | None | No |
| **Ancestry Family Tree Builder** | Yes | The ability to navigate a webpage, operate dropdown menus, drag-and-drop elements, and use the mouse and keyboard | Yes |
| **Chatbot** | No (but this study could change that) | The ability to use the mouse and keyboard | Yes |

## 2. RELATED WORK

### 2.1 Gathering Family Histories

The information-gathering process utilised by genealogists and family history researchers, FHRs for short, has been studied by multiple researchers in recent years. One such group of researchers, Duff and Johnson, suggest that the process utilised by FHRs comprises three phases: "tracing relevant ancestors; collecting relevant personal information; and gathering contextual information about their lives" [2, 3].

There are many ways to collect relevant personal information. The first step of this data retrieval process usually begins with historical records, such as the national archives and records administration or the state archives. Another common way to collect personal data when written records are lacking can be achieved through family history interviews. Thus far, these have been carried out in one of two ways. The first approach is via surveys that seek to capture specific pieces of information in a structured way. However, history has shown that this particular method is nowhere near as effective as one may think. Let's take Burke's Peerage as an example. Burke's Peerage is a genealogical publisher which focuses on keeping records on the ancestry of the peerage, baronetage, knightage and landed gentry of both Great Britain and Ireland. They utilise surveys in order to contact families in an attempt to try to gather their information. However, this approach has been seen to be very ineffective, as any data gatherer using surveys runs the risk of causing "survey fatigue", which can typically be separated into two types. The first type is "survey response fatigue", which occurs when people refrain from partaking in a survey before they even begin. This effect is normally a result of peoples' adherence to online and mail-based surveys and appears to be more prevalent with the elderly. The second type of fatigue is "survey-taking fatigue". It has been shown that the longer a survey takes, the less time and effort a person will put into it; their answers will be far less detailed and filled out much less rigorously [7]. Thus, surveys, in general, would be very ill-advised for family information gathering.

The second and evidently more effective solution is an in-depth interview, either face-to-face or by video call. These seek to capture as much information as possible over a long period of time through natural conversation [1]. It should be noted that this approach can be hindered by certain things, such as geographical constraints, e.g., if the data retriever is in Ireland and the person that they wish to interview is based in Alaska. This is where a conversational AI could be used to bridge this gap. Even though the amount of information retrieved may not be as complete as a human-to-human interaction, it could have greater potential than a regular survey.

### 2.2 Implementation

A chatbot, short for chatterbot or chat robot, is a piece of software that interacts with a user's queries through what is usually a text-based conversation. Its goal can range from providing simple pieces of information to a user to much more while simultaneously replicating the flow of a human-to-human conversation. The implementation of a chatbot to serve as a service agent to interact with during interview-like sessions has become somewhat popular in recent years, as it has been shown to be much more engaging and effective than a survey in an online setting [4, 6]. One such example is the "Wakamola" chatbot, which was developed to gather linked data to study the causes of obesity [4]. This chatbot employed multiple surveys based on standardised questionnaires, focusing on the participant's diets and physical activity levels [4]. This study, conducted with 61 volunteer students with a mean age of 20.5 years, showed that about half of the participants indicated an acceptable level of usability with such a chatbot [4].

Another approach is to employ the use of "service scripts" [6], which define the narrative that the chatbot will follow during its interactions. However, in an experimental scenario developed by Sands et al., some interesting findings were reported about the differences in opinion among 262 US consumers when dealing with a frontline service employee (FSE) or a script-based chatbot. Interestingly, with regard to the interaction effect, it was found that a service interaction with an FSE enhanced both rapport and user satisfaction and effectively outperformed a chatbot [6]. Though this may hint at an interaction with an FSE being preferable to an interaction with a chatbot, the use of an "entertaining service script" by a chatbot was shown to have a "marginally significant effect" on users with regard to rapport [6].

A user's engagement appears to be one of the most important aspects that need to be considered when designing a chatbot. Having a game-like experience by personifying the chatbot, giving it response feedback, and conversational follow-up, in order to keep the interaction going in a fluid and human-like way [4, 7], mixed with an appealing GUI, have been shown to be the most engaging and effective features of successful chatbots [4].

Rasa is an open-source conversational AI platform that allows users to create a conversational AI that can hold a conversation and connect to messaging channels and third-party systems, such as Facebook Messenger, through a set of APIs. In order to function properly, Rasa is based on the Python programming language and requires existing knowledge in both natural language processing (NLP) and chatbots in general. Rasa utilises YAML, a language popularly used for data serialisation, to manage all of its data, whether it be natural language understanding (NLU) data, stories or rules [11].

As for what each type of data is used for: the NLU data consists of examples of utterances categorised by intent. These are structured pieces of text extracted from a user's messages. It is also possible to add additional information, such as lookup tables, to help the model identify intents and entities correctly. The stories are used to train the model to recognise patterns, follow up on conversations, and to make it able to apply them to unseen interactions. Finally, the rules are also a representation of conversions between a user and a conversational AI. It describes small pieces of conversation that must always lead to the same flow and are used to train the "Rule Policy", which handles interactions with a fixed result and makes predictions from said rules [11].

For all of this to work, Rasa makes use of multiple natural language processing (NLP) and machine learning (ML) models in its API to facilitate the aforementioned script-based approach. Its core component is a recurrent neural network (RNN), which maps a history of raw dialogs into a distribution over system actions. It does so automatically, which reduces the amount of work for the developer with the manual feature engineering of dialog states [12]. To that core component, the previously mentioned rules and stories are added. Interestingly, Rasa already possesses some pre-trained neural networks, although none appear to be trained to handle genealogical conversations, as it seems to be a relatively untouched field.

One example of Rasa in use in a commercial setting is with T-Mobile. In this case, Rasa was tasked with building a new chatbot assistant to assist in customer interactions. This chatbot was able to customise the experience of its users by acting as a wayfinder or a transaction agent. Furthermore, although Rasa provided the company with a pre-trained and ready-to-go chatbot, T-mobile's team later personalised its virtual assistant and is still updating and improving on it [13].

Depending on the licence, paid or free, Rasa offers integration with both messaging platforms and with personal websites via the "Rasa Chat Widget" functionality. Though, it should be noted that all hosting must be done on a personal server, should one wish to avail of the free licence.

2.3 User Experience

The user experience provided by a chatbot is one of, if not the most important, features that should be considered during the design and development of a chatbot. In fact, recent studies have found that not only are the qualities of empathy and friendliness ones that a chatbot must display in order to build a level of trust between the user and itself, but the chatbot's perceived empathy had a greater impact on building trust with the user than did friendliness [9]. This same study also found that the complexity of the tasks provided to the user by a chatbot had a negative impact on the user's perceptions of a chatbot with respect to the aforementioned qualities [9]. In order to successfully gather information, a chatbot should be able to imitate natural conversation as much as "humanly" possible.

An important aspect of using a chatbot to consider during the design phase is "joy of use". It is incredibly important that the user enjoy themselves whilst conversing with a chatbot. Otherwise, the quality of their answers may suffer due to fatigue from use. This quality only becomes more vital in the context of family history interviews, as they often last a very long time and are most often with the elderly. It is of paramount importance that a genealogist be able to keep a conversation flowing naturally and in an enjoyable manner in order to avoid tiring out the interviewee. Any chatbot attempting to take on the role of a genealogist in this context should be able to do the same.

Some studies have suggested that an important balance between pragmatic quality (usefulness and efficiency) and hedonic quality ("joy of use") must be found [10] to create an effective chatbot, as though a high level of pragmatic quality may ensure that the chatbot's goals are met, a high level of hedonic quality may "benefit a user's well-being through engagement and stimulation" [10]. One such study was conducted by Haugeland et al., in which a randomised experiment was conducted with 35 participants to ascertain whether or not the use of topic-led conversations and free text interaction impacted a user's impressions of a chatbot [10]. In this study, the impact of the type of conversation on a user's experience was measured, and it was found that the levels of perceived anthropomorphism and hedonic quality were benefited by topic-led conversations, as some users reported that they appreciated a sense of progress or achievement in such conversations [10]. On the other hand, investigations into free-text approaches yielded results indicating that human likeness "may not depend primarily on free text interaction, but rather on high levels of flexibility and adaptivity". [10].

## 3. METHODOLOGY

### 3.1 Chatbot Implementation

In order to minimise the amount of time needed to develop a functioning chatbot to suit our purposes, a third-party, open-source, Python-based conversational AI platform named Rasa [14] was utilised. This platform provided a simple way for us to define a natural language understanding (NLU) pipeline in YAML to which we could add additional components, such as additional featurisers, entity extractors, or pre-trained language models [15] (see Figure 1 for an overview of Rasa's NLU pipeline).

Some additional optimisations were also made to increase the chatbot's ability to correctly identify and extract dates, locations, and people's names. These optimisations were two-fold: firstly, Spacy, a publicly available collection of pre-trained language models [16], was added to our existing NLU pipeline; and secondly, a collection of lookup tables were either manually created or sourced from a variety of locations to provide additional training data on which the Spacy models may not have been trained, such as French or Irish surnames, or lesser known towns and cities across the world, to name a few. Once a console-based minimum viable product (MVP) of this chatbot was complete, it was hosted as a widget within a Flask-based single-page application (SPA) to make for easier access and to allow for it to interact with our graphical companion (see 3.2 Graphical Companion). A local database, powered by SQLite, was also implemented in order to make all information learned about the user and their ancestors easily accessible to said graphical companion.

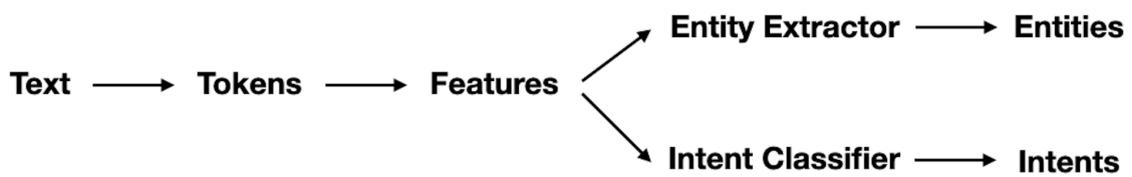

Figure 1. Rasa's NLU pipeline [15].

3.2 Graphical Companion

In order to allow the users to see their family trees as they were being constructed, a graphical companion was also implemented and embedded in our single-page application. This graphical companion, developed using Javascript, with the aid of a third-party library called "js_family_tree" [18], presented itself as a large window to the left of the chat widget in which a family tree was displayed. This family tree was updated using Javascript whenever a new piece of information was learned about either the user or one of their ancestors. The aim here was to not only illustrate to the user that the chatbot was taking what they were saying on board and learning but also to provide a "progress bar" of sorts in the form of a tree to assure them that they were making steps in the right direction. We believe that this was a worthwhile addition, as without any graphical feedback, the user may have felt that they were talking to something that was not actually understanding and making note of the information that they were providing. See Figure 2 for an example of a tree generated using "js_family_tree".

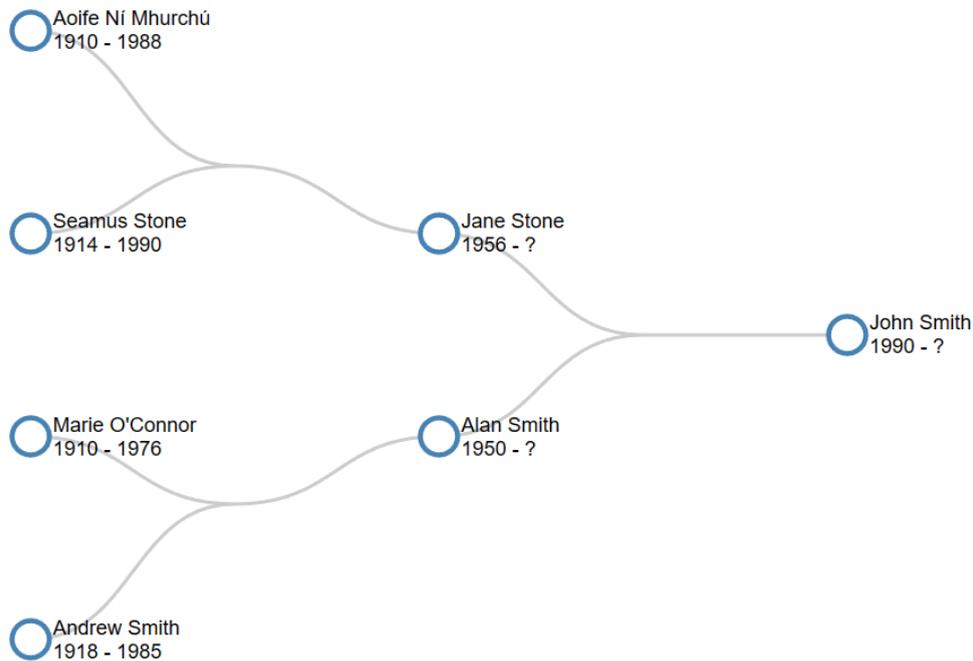

Figure 2. An example of a tree that has been generated using "js_family_tree".

## 4. EXPERIMENTS & RESULTS

### 4.1 Sessions Conducted

In order to perform a proper comparison between chatbots, in-person interviews, and ancestry.com's family tree builder, three separate sessions were conducted with each participant and were as follows:

**Ancestry family tree builder** - this set of sessions involved each participant in the study being asked to construct their own family tree using the family tree builder currently provided on ancestry.com.

**Chatbot** - this set of sessions involved each participant in the study being asked to construct their own family tree using our own chatbot.

**In-person interviews** - this was a series of in-person interviews, in which we played the role of the interviewer and attempted to learn enough to construct a family tree for the participants on paper.

It should be noted that for any sessions involving the use of technology, the participant was asked to carry out their task without any external aid in order to ensure that all tests were done fairly. Additionally, all sessions were seen to be complete when the information gathered was sufficient to construct a family tree up to and including the participant's grandparents.

### 4.2 Participants

For the benefit of this study, an effort was made to find the oldest participants possible due to the increased likelihood of them knowing more about their family histories. With this in mind, a group of six participants were chosen, with ages ranging from 49 to 90, an average age of 60, differing mother tongues: English and French, and differing levels of technical ability, from those who were completely unfamiliar with how to use a mouse and keyboard to those who were somewhat technically proficient (able to operate a computer, browse the internet, send emails, etc.). In the case of the participants who spoke French, the host of the sessions interpreted on their behalf.

### 4.3 Evaluation Metrics

With the varying nature of the sessions conducted in mind, three metrics were chosen to quantitatively measure the performance of each method tested; two, in the case of the in-person interviews, as one of the metrics devised was not applicable. They were as follows:

**Time taken** - this was the time taken for the participant or host, in the case of the in-person interviews, to construct a family tree up to and including their grandparents. This was measured in minutes and seconds only and was defined as the time from when the user began interacting with the platform being tested or when the first question was asked, in the case of the in-person interviews, to when both the host and participant were satisfied that sufficient information had been gathered to construct a family tree.

**Number of pauses due to confusion -** this was the number of times that the user needed to stop and seek guidance due to some form of confusion with the UI or the overall process. This was used as a measurement for both the chatbot and Ancestry sessions but was omitted from the in-person interviews as it would not have been a viable measurement.

**Number of corrections made** - this was the number of corrections that needed to be made by the user. In the case of both the chatbot and Ancestry sessions, this was counted as any time when the user needed to correct a misunderstanding on the software's part or a typo, whereas, with the in-person interviews, this was any time that the host needed to clarify some information or amend something due to a misunderstanding. We believe that these three metrics were sufficient to gauge the viability and usability of the three methods from a totally unbiased point of view.

### 4.4 Results

All sessions conducted followed the same steps based on a predefined plan to ensure consistency.

#### 4.4.1 In-Person Interviews

For all interviewees, the time taken to conclude the interview remained within a similar range, with an average time of 7:14. The average number of corrections required due to a misunderstanding was recorded as 0.83 per session. Though the age difference does not appear to have had much impact in an in-person interview regarding the number of corrections needed, said age does appear to have affected the time taken (see Table 2). Our conclusion here is that this was likely due to the answers from the older participants being less to the point and more extended.

Table 2. Results from the In-person Interviews.

|               | Age of participant | Time taken | Number of corrections made |
|---------------|--------------------|------------|----------------------------|
| Participant 1 | 49                 | 6:13       | 0                          |
| Participant 2 | 52                 | 6:01       | 1                          |
| Participant 3 | 52                 | 5:03       | 0                          |
| Participant 4 | 58                 | 6:46       | 1                          |
| Participant 5 | 58                 | 8:44       | 0                          |
| Participant 6 | 90                 | 10:40      | 3                          |

4.4.2 Ancestry's Family Tree Builder

Unlike with the in-person interviews, the Ancestry session appeared to highlight some differences between participants with low or high levels of technical proficiency. The time taken for participants to construct a family tree on ancestry.com ranged from 4:50 to 27:09, with an average time taken of 15:36, an average of 1.83 corrections required due to typos or misunderstandings, and an average of 3.5 pauses due to confusion with the UI or next steps required (see Table 3). A common error that was observed during the sessions was with the automatic setting of an ancestor's maiden name, which, at times, was incorrect and briefly caused confusion.

Table 3. Results from the Ancestry Family Tree Builder Sessions.

|               | Age of participant | Time taken | Number of corrections made | Number of pauses due to confusion |
|---------------|--------------------|------------|----------------------------|-----------------------------------|
| Participant 1 | 49                 | 14:56      | 1                          | 3                                 |
| Participant 2 | 52                 | 4:50       | 1                          | 2                                 |
| Participant 3 | 52                 | 11:46      | 3                          | 5                                 |
| Participant 4 | 58                 | 27:09      | 1                          | 2                                 |
| Participant 5 | 58                 | 25:40      | 2                          | 6                                 |
| Participant 6 | 90                 | 9:19       | 3                          | 3                                 |

4.4.3 Chatbot

Similarly to the Ancestry sessions, the chatbot sessions also showed differing outcomes based on both the ages and technical proficiencies of the participants. The time taken for the participant to construct a family tree by speaking to our chatbot was somewhat longer than the previous session, ranging from 9:50 to 36:52 with an average time taken of 22:24. With regards to the number of errors caused by the software, more corrections were required in comparison to the Ancestry session, with an average of 4.17 corrections made. However, the chatbot did appear to cause less confusion than the Ancestry family tree builder, with an average of 2.5 pauses required per session (see Table 4). In these sessions, interviewees from the "older" generations, although slower, seemed to get less confused and caused fewer errors than the

"younger" ones. Indeed, the chatbot's way of working appeared to be more compatible with an elderly participant's way of thinking and answering. Where the younger interviewees tended to give short answers mainly structured with simple keywords, the less experienced gave full sentences, which were much more easily interpreted by the chatbot.

Table 4. Results from the Chatbot Sessions.

|  | Age of participant | Time taken | Number of corrections made | Number of pauses due to confusion |
|---|---|---|---|---|
| Participant 1 | 49 | 19:25 | 4 | 1 |
| Participant 2 | 52 | 9:50 | 7 | 4 |
| Participant 3 | 52 | 18:10 | 6 | 5 |
| Participant 4 | 58 | 36:52 | 2 | 3 |
| Participant 5 | 58 | 29:55 | 5 | 2 |
| Participant 6 | 90 | 20:16 | 1 | 0 |

4.4.4 Overview

Overall, the in-person interviews demonstrated the fastest completion times and the highest accuracy. The Ancestry and chatbot sessions yielded comparable results in terms of completion time but with slightly higher error rates in the case of the chatbot. It is worth noting that the chatbot interviews showed promise in terms of usability and convenience, but further refinement is required to reduce the error rate and improve the accuracy of responses. Also, even though this metric wasn't taken into account as it is neither objective nor quantifiable, a general opinion was that the chatbot was more enjoyable to work with than the tree builder. See Table 5 for an overview of the aforementioned results.

Table 5. Overall Result.

| Interview Method | Average Time taken | Average Number of corrections made | Average Number of pauses due to confusion |
|---|---|---|---|
| **In-person** | 7:14 | 0.83 | N/A |
| **Ancestry** | 15:36 | 1.83 | 3.5 |
| **Chatbot** | 22:24 | 4.17 | 2.5 |

5. DISCUSSION

5.1 Findings

Our research has shown that an in-person interview was more effective than a chatbot conversation. However, the chatbot was still able to achieve results comparable to other existing means of gathering family history information. In the context of family history retrieval, this study has shown with a decent degree of confidence that using a conversational AI to gather family history information is a viable alternative to the more regularly employed questionnaire approach [4, 6].

5.2 Limitations & Weaknesses

With regard to limitations in our research, several aspects of our implementation and experimentation should be mentioned. Though the chatbot used a pre-trained model for entity extraction, it still struggled at times with certain names, both people's names and locations. This indicates a need for additional training with a larger and more varied training set to increase the chatbot's ability to extract these names from user input. In the case of the Rasa platform, in retrospect, though it provided a great deal of out-of-the-box functionality, alternative platforms may be worth considering, as our use of Rasa has highlighted certain caveats that would indicate that said platform is catered more towards the implementation of virtual assistants, rather than conversation AI. The conversations are not led by the chatbot directly, rather, it has to infer what to say next from previous user interactions predefined stories. Lastly, although we believe that the size of our pool of participants was sufficient to test out the viability of using a conversational AI in the place of an in-person interview, a larger pool would have been preferable from a user-testing point of view.

5.3 Future Considerations

Looking ahead, future research could focus on enhancing the performance of our chatbot by refining and adding to the training data used. Additionally, using an alternative framework to build an entirely new proof of concept could be viable, provided that a platform exists that can overcome some of the observed issues with Rasa. Finally, conducting studies with larger and more diverse samples could also provide a more comprehensive understanding of user engagement and the broader applicability of virtual conversation agents for FHR.

5.4 Conclusion

In conclusion, although our research has shown the value of in-person interviews for efficient data collection with minimal errors, we have highlighted how automated approaches like chatbots offer viable alternatives, showcasing their usability and convenience. Based on previous literature, it would appear, from our previous searches, that a chatbot has not previously been used in this context. Such a potentially novel approach could provide a useful alternative to in-person interviews where geographical constraints are in play, such as when one's desired interviewees may live in another country that is not easily accessible. In addition, compared to existing means of gathering family history information, such as questionnaires, a chatbot-based approach could invoke more engagement from the desired interviewees who, due to their age, are more often than not uncomfortable with technology. With further refinement, chatbots have the potential to become valuable tools for genealogists with regard to the retrieval of family history information.